\documentclass{article}

\usepackage{microtype}
\usepackage{graphicx}
\usepackage{booktabs} 

\usepackage[mode=buildnew]{standalone}
\usepackage{lipsum}

\usepackage{amsmath}

\usepackage{algorithm}
\usepackage{algorithmic}

\usepackage{multirow}
\usepackage{enumitem}

\usepackage{graphicx}
\usepackage{subcaption}

\usepackage{fancybox}

\usepackage{hyperref}

\usepackage[accepted]{mlsys2024}

\mlsystitlerunning{Balancing Pipeline Parallelism with Vocabulary Parallelism}

\usepackage{listings}
\usepackage{xcolor} 

\definecolor{background}{RGB}{245, 245, 245}
\definecolor{comments}{RGB}{0, 128, 0} 
\definecolor{keywords}{RGB}{0, 0, 255} 
\definecolor{strings}{RGB}{163, 21, 21} 
\definecolor{numbers}{RGB}{128, 0, 128} 

\lstdefinestyle{bashstyle}{
    backgroundcolor=\color{background}, 
    basicstyle=\ttfamily\small, 
    commentstyle=\color{comments}, 
    keywordstyle=\color{keywords}, 
    stringstyle=\color{strings}, 
    numberstyle=\color{numbers}, 
    breakatwhitespace=false, 
    breaklines=true, 
    captionpos=b, 
    keepspaces=true, 
    numbers=none, 
    showspaces=false, 
    showstringspaces=false, 
    showtabs=false, 
    tabsize=2, 
    frame=single, 
    framesep=5pt, 
    rulecolor=\color{black}, 
    language=bash, 
    morekeywords={sudo, apt, echo, if, then, else, fi, for, do, done, while, case, esac, function}, 
    morecomment=[l]{\#}, 
    morestring=[b]{"}, 
    morestring=[b]{'}, 
}

\usepackage{graphicx}
\usepackage{hyperref}
\usepackage{eso-pic}
\usepackage{xparse}
\usepackage{etoolbox}

\newlength{\badgewidth}
\newlength{\badgegap}
\newcommand{\badgeList}{}

\NewDocumentCommand{\addTopRightBadge}{O{} m}{%
\gappto{\badgeList}{\href{#1}{\includegraphics[width=\badgewidth]{#2}}\hspace{\badgegap}}%
}

\newcommand{\placeTopRightBadges}{%
\AddToShipoutPictureBG*{%
\put(\LenToUnit{\paperwidth - 1.5cm - \badgewidth},\LenToUnit{\paperheight - 2cm}){%
\makebox[0pt][r]{\badgeList}%
}%
}%
}

\addTopRightBadge{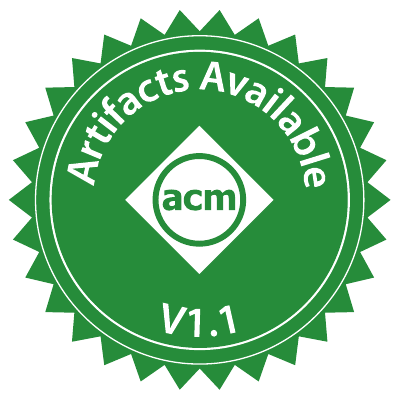}

\placeTopRightBadges

\begin{document}

\twocolumn[
\mlsystitle{Balancing Pipeline Parallelism with Vocabulary Parallelism}



\mlsyssetsymbol{equal}{*}
\mlsyssetsymbol{intern}{\dag}

\begin{mlsysauthorlist}
\mlsysauthor{Man Tsung Yeung}{intern,sea,nus}
\mlsysauthor{Penghui Qi}{sea,nus}
\mlsysauthor{Min Lin}{sea}
\mlsysauthor{Xinyi Wan}{sea}
\end{mlsysauthorlist}

\mlsysaffiliation{sea}{Sea AI Lab}
\mlsysaffiliation{nus}{National University of Singapore}

\mlsyscorrespondingauthor{Xinyi Wan}{wanxy@sea.com}

\mlsyskeywords{Machine Learning, MLSys}

\vskip 0.3in

\begin{abstract}

Pipeline parallelism is widely used to scale the training of transformer-based large language models, various works have been done to improve its throughput and memory footprint. In this paper, we address a frequently overlooked issue: the vocabulary layers can cause imbalanced computation and memory usage across pipeline stages, worsening pipeline bubbles and the memory bottleneck. To tackle this, we partition the vocabulary layers evenly across pipeline devices and group the computation into pipeline passes. To reduce the activation memory overhead, we propose several algorithms to reduce communication barriers within vocabulary layers. Additionally, we utilize a generalizable method to integrate Vocabulary Parallelism with existing pipeline schedules. By combining these techniques, our methods effectively balance the computation and parameter memory, with only a small constant activation memory overhead. Notably, when combined with activation memory-balanced schedules like \textit{V-Half}, our approach achieves perfect balance in both memory and computation. Extensive evaluations demonstrate that our method achieves computation and memory balance regardless of the vocabulary size, resulting in a 5\% to 51\% improvement in throughput compared to na\"ive approaches, meanwhile significantly reducing peak memory usage especially for large vocabulary scenarios. Our implementation is open-sourced at \href{https://github.com/sail-sg/VocabularyParallelism}{https://github.com/sail-sg/VocabularyParallelism}.
\end{abstract}
]



\printAffiliationsAndNotice{\textsuperscript{\dag} Work was done during an internship at Sea AI Lab.}  

\section{Introduction}

As the scale of transformer models \cite{vaswani2017attention, brown2020language} continues to grow, model parallelism has garnered significant attention within the deep learning community. Several model parallel techniques have been proposed to address the challenges associated with training large models, including Zero Redundancy Optimizer (ZeRO)~\citep{rajbhandari2020zero, zhao2023pytorch}, Tensor Parallelism (TP)~\citep{shoeybi2019megatron}, and Pipeline Parallelism (PP)~\cite{huang2019gpipe, 1f1b, megatron-lm, zb, controlable-mem}. Each of these methods has its own advantages and limitations. For instance, ZeRO is effective in reducing memory by eliminating redundant parameter storage, but suffers from high communication overhead when gathering partitioned parameters and gradients for scenarios with limited network bandwidth or requiring frequent parameter updates. TP can efficiently handle large model parameters by splitting them across devices, but often faces low arithmetic intensity and requires significant inter-device communication.
Among these techniques, PP shows distinct advantages due to its low communication cost and high arithmetic intensity, making it particularly attractive for training large-scale models. However, PP faces two significant challenges: pipeline bubbles and high memory consumption. Pipeline bubbles occur when there are idle periods in the pipeline stages, leading to suboptimal utilization of computational resources. Various strategies have been proposed to mitigate pipeline bubbles, such as token-level PP~\citep{li2021terapipe} and interleaved 1F1B~\citep{megatron-lm}. An exceptional advancement is zero-bubble pipeline parallelism~\citep{zb, controlable-mem}, which achieves almost zero bubble in many scenarios through splitting backward pass into activation gradient computation and weight gradient computation.
In most PP schedules, the activations of several microbatches are stored to reduce pipeline bubbles, making memory a critical bottleneck to scale large models. Previous work has explored activation recomputation~\citep{chen2016training, korthikanti2023reducing}, memory transferring~\cite{kim2023bpipe} and memory-efficient V-Shape scheduling~\citep{controlable-mem} to mitigate this issue. Despite various effort, the memory bottleneck still poses the largest limitation for PP.

\begin{figure}
    \centering
    \includegraphics[width=\linewidth]{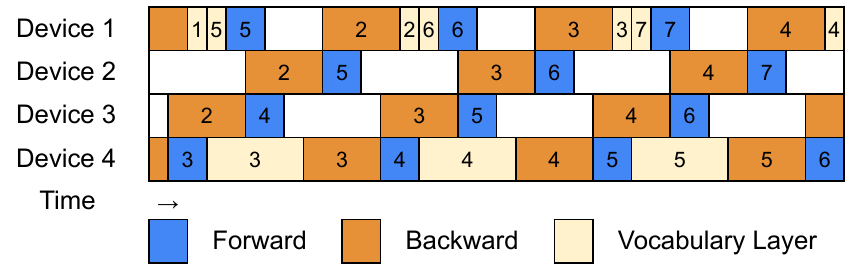}
    \caption{Repeating pattern in an imbalanced pipeline. Bubbles are incurred due to an extra output layer in the last pipeline stage.}
    \label{fig:imbalanced-stable}
\end{figure}

\begin{figure}[h]
    \centering
    \includegraphics[width=\linewidth]{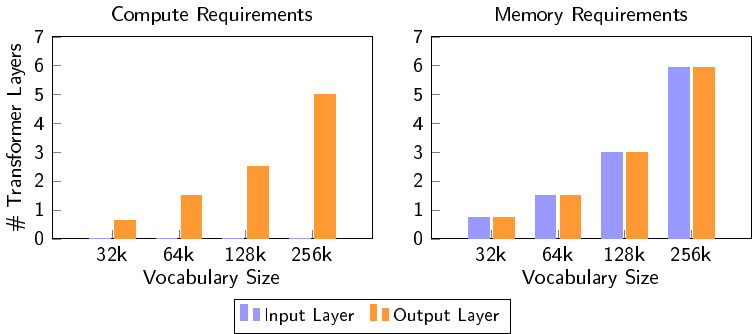}
    \caption{Ratio of compute and memory of vocabulary layers compared to transformer layers in Gemma2-9B.}
    \label{fig:vocab-layers}
\end{figure}

In this paper, we focus on an imbalance issue in PP caused by vocabulary-related layers, which is often overlooked in practice but can significantly degrade the performance in both throughput and memory. Typically, transformer layers are uniformly distributed across pipeline stages, while the first stage contains an additional input layer and the last stage contains an additional output layer. Such imbalanced setup greatly hurts the performance in both computation and memory. Firstly, pipeline bubbles are introduced in other pipeline stages due to their less workload, as shown in Figure \ref{fig:imbalanced-stable}. Additionally, the additional input layer in the first stage exacerbates the memory bottleneck for most PP schedules like 1F1B~\citep{1f1b}. Finally, as the vocabulary size grows larger \citep{tao2024scaling}, this imbalance becomes more pronounced, as shown in Figure \ref{fig:vocab-layers}. For instance, in the case of Gemma2 9B with a vocabulary size of 256k \citep{team2024gemma}, both the computation and parameter memory of the output layer are approximately 5 times those of the transformer layers, highlighting the severity of the issue.

To address this imbalance issue, we propose a novel Vocabulary Parallelism approach to balance the computation and memory in PP. By partitioning the vocabulary layers across all pipeline devices, we introduce several methods to group the computation and communication barriers together with a generalizable scheduling approach in PP, with only a small constant memory overhead. Extensive experiments demonstrate our approach significantly outperforms na\"ive layer redistribution and other existing methods, resulting in up to 51\% improvement in throughput.

\section{Related Work}

\paragraph{Balancing Activation Memory}
A line of research addresses another aspect of imbalance in PP, the activation memory with the 1F1B schedule. For instance, BPipe~\citep{kim2023bpipe} transfers activations between devices, trading communication for reduced peak memory. Another approach uses V-Shape scheduling to create a pipeline schedule with balanced and efficient memory usage. These methods are orthogonal to our work, and combining them can achieve fully balanced pipeline parallelism in both computation and memory (activations and parameters).



\begin{figure}[t]
    \centering
    \includegraphics[width=\linewidth]{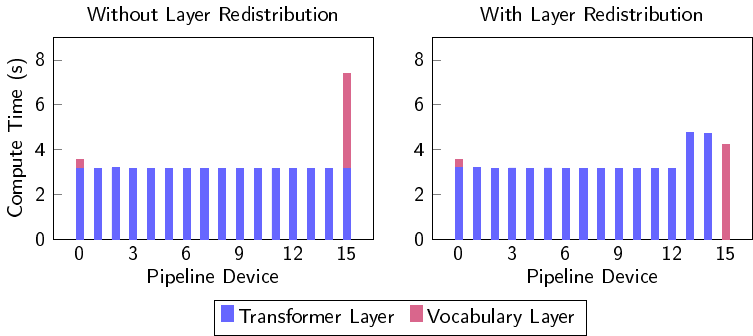}\\
    \vspace{2mm}
    \includegraphics[width=\linewidth]{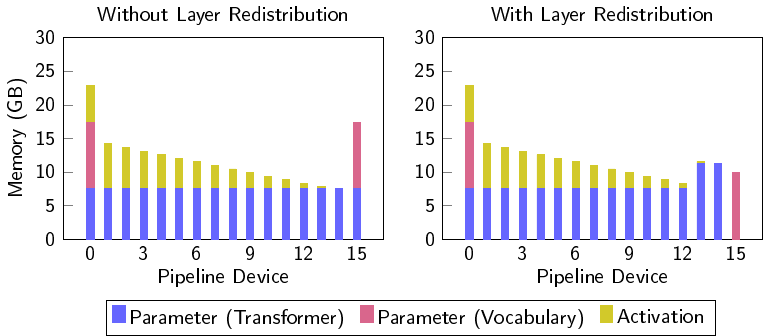}\\
    \caption{Transformer Layer Redistribution for a 7B GPT-like model with vocabulary size 128k. In this case, each stage has 2 transformer layers, while output layer is equivalent to 2.4x of transformer layer on compute and 2.6x on parameter memory.}
    \label{fig:redis}
\end{figure}

\paragraph{Balancing Vocabulary Layers}
Some existing training systems try to mitigate the imbalance caused by vocabulary layers by redistributing transformer layers across different stages. DeepSpeed \citep{deepspeedmegatron} uses a greedy method to automatically re-balance the workload between stages at the layer level. Similar strategies are employed in the training of Skywork-MoE \citep{wei2024skywork}. However, simply redistributing transformer layers faces several disadvantages. Firstly, even after redistribution, compute imbalance can still persist since only a subset of pipeline stages receive additional layers. An example is shown in Figure \ref{fig:redis}. This is particularly evident when the number of layers on each stage is small. Secondly, different layer types have varying compute-to-memory ratios, meaning that the re-balancing can only be based on either compute or memory but not both. In practice, the re-balancing is typically performed based on compute, leaving the memory imbalance still significant, particularly for input vocabulary layers that require minimal compute but substantial memory. Lastly, the effectiveness and planning of redistribution heavily depend on both the model settings and pipeline parallel settings. This makes it less flexible and challenging to adopt in various scenarios.

It's also worth noting that some models pretrained from scratch like Llama 3 \citep{dubey2024llama} reduce one transformer layer each from the first and the last stages, respectively. This method requires changes to architecture of models, which is out of the scope of this paper. Also, it has limitations if the training starts from a checkpoint where the number of transformer layers of model is fixed.

Another method to mitigate the imbalance problem in pipeline parallelism (PP) is the interlaced pipeline, reported by the automatic parallelism framework nnScaler~\citep{nnScaler}. This approach distributes the input and output vocabulary layers across different pipeline devices using a tensor parallel (TP) style \cite{megatron-lm}. By alternating between TP for vocabulary layers and PP for transformer layers, it aims to balance compute and memory overhead. However, TP requires frequent synchronization between devices, leading to two major drawbacks. First, the peak activation memory for 1F1B increases to 1.5 times of its original value (see Appendix \ref{app:interlaced_pipeline_memory}), which may make the critical memory bottleneck even worse. Second, the synchronized all-reduce during the output vocabulary layer introduces additional pipeline bubbles for each microbatch. Our ablation study in Appendix \ref{app:interlaced_pipeline_sync} shows these all-reduce along slows down the end to end training by approximately 11\% on 32 GPUs. These significant overheads in both activation memory and pipeline bubbles render the interlaced pipeline impractical in real-world scenarios.

\section{Vocabulary Parallelism in Pipeline Parallelism}

To completely address the imbalance issue in PP, we propose Vocabulary Parallelism under the following design principles:

\begin{itemize}
    \setlength{\itemsep}{0pt}
    \item We partition the vocabulary layers across the vocabulary dimension, and distribute it evenly to all pipeline devices.
    \item To be native to pipeline parallelism, the computation of vocabulary layers should be represented as passes similar to forward/backward passes of transformer layers.
    \item Integrating vocabulary passes into pipeline schedules should not drastically affect the memory and efficiency of the original pipeline schedule.
\end{itemize}

Intuitively, after partitioning the vocabulary layers to all pipeline devices, computations on each device can be scheduled independently by inserting them cleverly into the existing pipeline passes, as long as the dependencies are still satisfied. However, it is worth noting that partitioning the softmax computation creates several all-reduce operations. These communication barriers create cross-device dependencies.

Driven by this intuition, in Section \ref{section:sync}, we discuss how to partition the computation in vocabulary layers to multiple devices and group them into pipeline passes. We observe that the communication barriers (e.g. the all-reduces in softmax) within the vocabulary layers increases the activation memory consumption of the pipeline schedule. As an improvement, we propose two novel algorithms to reduce the number of communication barriers, which reduces the activation memory overhead to minimum.

In Section \ref{section:scheduling}, we discuss how to integrate these vocabulary passes into existing pipeline schedules. Inspired by the framework presented in \citeauthor{controlable-mem} \yrcite{controlable-mem}, we insert the vocabulary passes into the building blocks of existing schedules and simply repeat building blocks to construct pipeline schedules. This relieves us from the hassle of deciding the ordering of vocabulary passes of every microbatch and is naturally generalizable to other schedules.

\begin{figure}
    \centering
    \includegraphics[width=\linewidth]{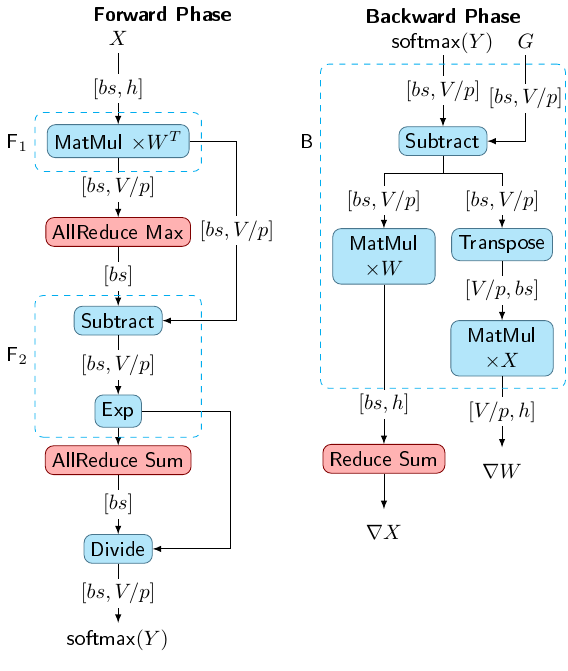}
    \caption{Computation graph of the output layer after partitioning across the vocabulary dimension. There are three all-reduce / reduce communications across all devices.}
    \label{fig:output-layer-graph}
\end{figure}


\section{Vocabulary Passes Construction} \label{section:sync}

In this section, we introduce how to split the vocabulary layers into several computation passes after partitioning them across all pipeline devices, and how to optimize the number of communication barriers.


\subsection{A Na\"ive Approach}

In the input layer, each device can perform forward and backward computations independently. We provide details on the input layer in Appendix \ref{appendix:input-layer}, and focus on the output layer for the remainder of this paper.

Figure \ref{fig:output-layer-graph} shows the computation graph of the output layer after partitioning the layer across $p$ devices. We denote the microbatch size as $b$, sequence length as $s$, hidden dimension as $h$ and vocabulary size as $V$.

\begin{figure}[H]
    \centering
    \includegraphics[width=\linewidth]{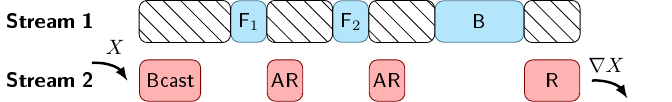}
    \caption{Overlapping all-reduce communication with transformer layer computation.}
    \label{fig:naive-streams}
\end{figure}

The partitioned output layer can be grouped into three computation passes $\textit{F}_1$, $\textit{F}_2$ and $\textit{B}$, separated by three all-reduce / reduce communications involving the maximum of logits, the sum of logits and the input gradients respectively. We can overlap these all-reduce communications with transformer layer computation by placing them in a separate stream, as shown in Figure \ref{fig:naive-streams}.

\begin{figure}[H]
    \centering
    \includegraphics[width=\linewidth]{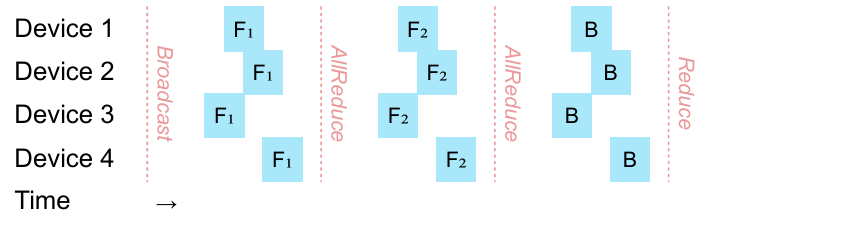}
    \caption{Scheduling dependencies in the na\"ive output layer implementation.}
    \label{fig:naive-dependencies}
\end{figure}

Figure \ref{fig:naive-dependencies} shows the computation and communication dependencies for a single microbatch. Notably, each of these all-reduce communications will introduce a communication barrier across all pipeline devices, which complicates the pipeline scheduling. As shown later in Section \ref{section:scheduling-examples}, the number of communication barriers also increases the activation memory consumption of the pipeline schedule. Therefore, we aim to reduce the number of communication barriers by reordering the operations in the output layer.

\subsection{Definitions}

We detail the computation in the output layer. Given the output of the last transformer layer $X$ and the embedding weights $W$, we first compute $Y$:
\begin{equation}
    Y = XW^T
\end{equation}

Then, the (safe) softmax of each sequence is computed as follows:
\begin{equation}
    \text{softmax}(Y_{ij}) = \dfrac{e^{Y_{ij} - m_i}}{\text{sum}_i} 
\end{equation}

where $m_i = \max_{k} Y_{ik}$ is the maximum of the logits and $\text{sum}_i = \sum_{k} e^{Y_{ik} - m_i}$ is the sum of logit exponents.

Assuming the cross entropy loss is used, in the backward phase, we have
\begin{align}
    \nabla X & = \left( \text{softmax}(Y) - G \right) W\\
    \nabla W & = \left( \text{softmax}(Y) - G \right)^T X
\end{align}

where $G$ is the ground truth matrix with $G_{ig_i} = 1$ and $G_{ij} = 0$ otherwise, where $g_i$ is label for token $i$.

\subsection{Forward Phase Optimization}

Inspired by online softmax \cite{online-softmax, dao2022flashattention}, we observe that the all-reduce communication for $m_i$ and $\text{sum}_i$ can be done after computing the softmax. Instead of using the global maximum and sum, each device instead computes $\text{softmax}'(Y_i)$ using the local maximum and sum from its vocabulary partition. We then have
\begin{align}
    \text{softmax}(Y_{ij}) = \text{softmax}'(Y_{ij}) \times \dfrac{\text{sum}'_i \times e^{m'_i - m_i}}{\text{sum}_i} \label{eq:softmax-fast}
\end{align}
where $m_i'$ and $\text{sum}'_i$ are the locally computed maximum and sum, respectively.

Using equation \ref{eq:softmax-fast}, we have Algorithm \ref{alg:async1} that reduces the 3 communication barriers to 2, which are denoted as $C_1$ and $C_2$ respectively.

\cornersize{.2}
\begin{algorithm}[H]
\caption{Output layer with 2 communication barriers}\label{alg:async1}
\begin{algorithmic}
\FUNCTION{forward\_and\_backward($W$)}
\STATE $X \gets$ Receive Broadcast \hfill $\triangleleft$ $C_0$
\STATE \ovalbox{\begin{minipage}{6cm}
    $Y \gets XW^T$\\
    $m'_i \gets \max_{k=1}^{V/p} Y_{ik}$\\
    $\text{sum}'_i \gets \sum_{k=1}^{V/p} e^{Y_{ik} - m'_i}$\\
    $\text{softmax}'(Y_{ij}) \gets \dfrac{e^{Y_{ij} - m'_i}}{\text{sum}'_i}$
\end{minipage}}\hfill $\triangleleft$ $S$
\STATE \ovalbox{\begin{minipage}{6cm}
    $m_i \gets$ AllReduce $m'_i$\\
    $\text{sum}'_i \gets \text{sum}'_i \times e^{m'_i - m_i}$\\
    $\text{sum}_i \gets $ AllReduce $\text{sum}'_i$
\end{minipage}}\hfill $\triangleleft$ $C_1$
\STATE \ovalbox{\begin{minipage}{6cm}
    $\text{softmax}(Y_i) \gets \text{softmax}'(Y_i) \times \dfrac{\text{sum}_i'}{\text{sum}_i}$\\
    $\nabla X' \gets$ $(\text{softmax}(Y) - G) W$\\
    $\nabla W \gets (\text{softmax}(Y) - G)^T X$
\end{minipage}}\hfill $\triangleleft$ $T$
\STATE $\nabla X \gets$ Reduce $\nabla X'$ \hfill $\triangleleft$ $C_2$
\ENDFUNCTION
\end{algorithmic}
\end{algorithm}

The elementwise operations in $C_1$ only involves tensors of size $[bs]$ as opposed to size $[bs, V/p]$, which greatly reduces the computation pressure when overlapped with transformer layer computation.

\subsection{Backward Phase Optimization}

We also observe that all three all-reduce / reduce communications can be done after computing the matrix multiplications for the input gradients. Note that
\begin{align}
    \nabla X = \text{softmax}'(Y) W \times \dfrac{\text{sum}'_i \times e^{m'_i - m_i}}{\text{sum}_i} - GW \label{eq:input-grad-fast}
\end{align}
We can compute $\text{softmax}'(Y) W$ and $GW$ beforehand, and all-reduce $\nabla X$ after we obtain $m_i$ and $\text{sum}_i$. Since the matrix multiplications in equation \ref{eq:input-grad-fast} are already computed, computing $\nabla X$ within the communication barrier only involves lightweight operations.

This allows us to complete both phases in the output layer with only a single communication barrier $C_1$, as shown in Algorithm \ref{alg:async2}.

\begin{algorithm}[H]
\caption{Output layer with 1 communication barrier}\label{alg:async2}
\begin{algorithmic}
\FUNCTION{forward\_and\_backward($W$)}
\STATE $X \gets$ Receive Broadcast \hfill $\triangleleft$ $C_0$
\STATE \ovalbox{\begin{minipage}{6cm}
    $Y \gets XW^T$\\
    $m'_i \gets \max_{k=1}^{V/p} Y_{ik}$\\
    $\text{sum}'_i \gets \sum_{k=1}^{V/p} e^{Y_{ik} - m'_i}$\\
    $\text{softmax}'(Y_{ij}) \gets \dfrac{e^{Y_{ij} - m'_i}}{\text{sum}'_i}$\\
    $A \gets \text{softmax}'(Y) W$\\
    $B \gets GW$
\end{minipage}}\hfill $\triangleleft$ $S$
\STATE \ovalbox{\begin{minipage}{6cm}
    $m_i \gets$ AllReduce $m'_i$\\
    $\text{sum}'_i \gets \text{sum}'_i \times e^{m'_i - m_i}$\\
    $\text{sum}_i \gets $ AllReduce $\text{sum}'_i$\\
    $\nabla X \gets $ Reduce $A \times \frac{\text{sum}'_i}{\text{sum}_i} - B$
\end{minipage}}\hfill $\triangleleft$ $C_1$
\STATE \ovalbox{\begin{minipage}{6cm}
    $\text{softmax}(Y) \gets \text{softmax}'(Y) \times \dfrac{\text{sum}_i'}{\text{sum}_i}$\\
    $\nabla W \gets (\text{softmax}(Y) - G)^T X$
\end{minipage}}\hfill $\triangleleft$ $T$
\ENDFUNCTION
\end{algorithmic}
\end{algorithm}

Note that the weight gradient step $T$ can be arbitrarily delayed since no other operations depend on it, similar to the idea in zero-bubble strategy~\cite{zb}.

\begin{figure}
    \centering
    \includegraphics[width=\linewidth]{figures/naive.pdf}\\
    $\downarrow$\\
    \vspace{1mm}
    \includegraphics[width=\linewidth]{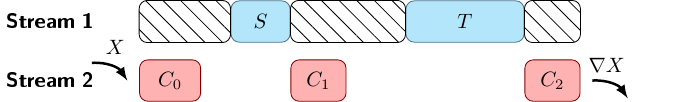}\\
    $\downarrow$\\
    \vspace{1mm}
    \includegraphics[width=\linewidth]{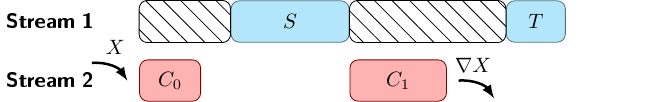}
    \caption{Computation order in the output layer for a single microbatch, corresponding to the na\"ive implementation, Algorithm \ref{alg:async1} and Algorithm \ref{alg:async2} respectively.}
    \label{fig:opt-computation-graph}
\end{figure}

We compare the two algorithms with the na\"ive implementation in Figure \ref{fig:opt-computation-graph}. By placing the operations in the communication barrier in a separate stream, they will be able to overlap with transformer layer computation. Compared to Algorithm \ref{alg:async1}, Algorithm \ref{alg:async2} introduces a bit more computation overhead (shown in Section \ref{section:overhead}). However, it is still beneficial to reduce the number of communication barriers. As shown in section \ref{section:scheduling-examples}, a reduction in the number of communication barriers will help to save activation memory.

\section{Pipeline Scheduling} \label{section:scheduling}

In this section, we show a systematic method about how to make minimal changes to typical pipeline schedules to include the output layer passes. We apply our method on two different schedules, 1F1B \cite{1f1b} and \textit{V-Half} \cite{controlable-mem}. Despite its popularity, an inherent problem of the 1F1B schedule is an imbalanced activation memory across pipeline devices. In contrast, \textit{V-Half} balances out the activation memory by a V-Shape device placement, reducing the activation memory requirement to half of that of 1F1B. By integrating Vocabulary Parallelism into \textit{V-Half}, we aim to achieve a completely memory-balanced pipeline.

\subsection{Scheduling Dependencies}

In Algorithms \ref{alg:async1} and \ref{alg:async2}, we perform output layer computation with 2 and 1 communication barriers, respectively. For each microbatch, we have to integrate the output layer passes, $S$ and $T$, into the pipeline schedules. The pipeline schedule has to adhere to the following constraints:

\begin{itemize}[topsep=0pt]
    \setlength{\itemsep}{0pt}
    \item All $S$ passes must be scheduled after the forward pass of the last transformer layer completes.
    \item All $T$ passes must be scheduled after all $S$ passes complete.
    \item For Algorithm \ref{alg:async1} only, the backward pass of the last transformer layer must be scheduled after all $T$ passes complete. In contrast, the $T$ passes can be arbitrarily delayed in Algorithm \ref{alg:async2}.
\end{itemize}

\begin{figure}[H]
    \centering
    \includegraphics[width=\linewidth]{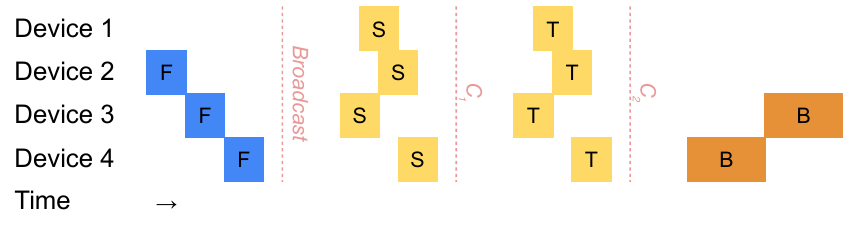}
    \includegraphics[width=\linewidth]{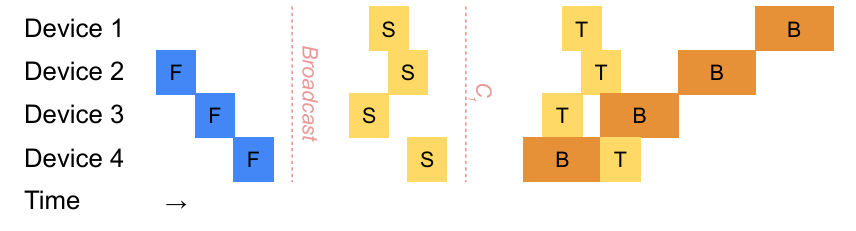}
    \caption{Scheduling Dependencies in Algorithms \ref{alg:async1} and \ref{alg:async2}.}
    \label{fig:scheduling-dependencies}
\end{figure}

For example, Figure \ref{fig:scheduling-dependencies} shows the scheduling dependencies for a single microbatch in Algorithms \ref{alg:async1} and \ref{alg:async2} respectively.

\subsection{Methodology} \label{section:scheduling-examples}

\begin{figure}[t]
    \centering
    \includegraphics[width=\linewidth]{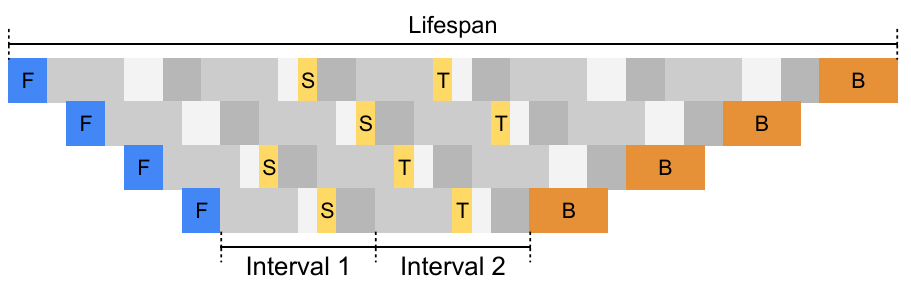}
    \includegraphics[width=\linewidth]{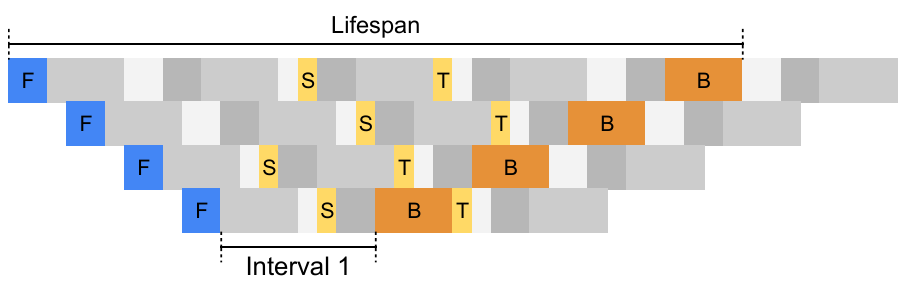}
    \caption{Modified building blocks for the 1F1B schedule corresponding to Algorithm \ref{alg:async1} and Algorithm \ref{alg:async2}. The output layer passes are inserted.}
    \label{fig:building-block}
\end{figure}

To elegantly integrate these $S$ and $T$ passes into typical pipeline schedules while adhering to the constraints, we follow \citeauthor{controlable-mem}'s framework \yrcite{controlable-mem} to construct pipeline schedules. In this framework, each pipeline schedule can be structured by its building block, which is defined by the scheduling pattern of each microbatch. By uniformly repeating a building block, we can construct a pipeline schedule, with peak activation memory calculated by dividing its lifespan by its interval. The lifespan is the time between a forward and its corresponding backward, while the interval is the workload of a single microbatch on each device, as illustrated in Figure \ref{fig:building-block}. This approach greatly simplifies dependency management for each microbatch and facilitates memory consumption analysis.

\begin{figure*}[t]
    \centering
    \includegraphics[width=\linewidth]{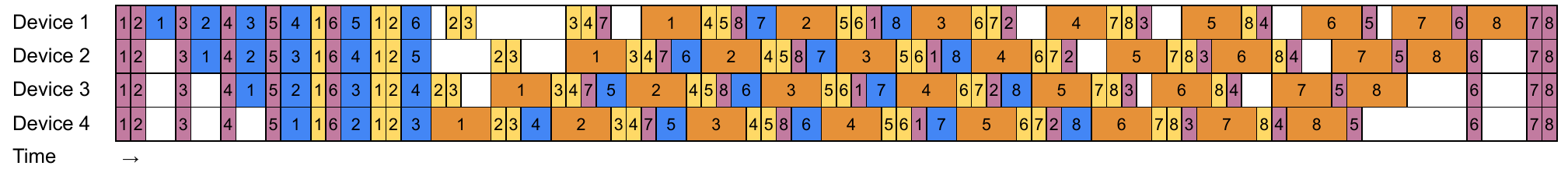}\\
    \includegraphics[width=\linewidth]{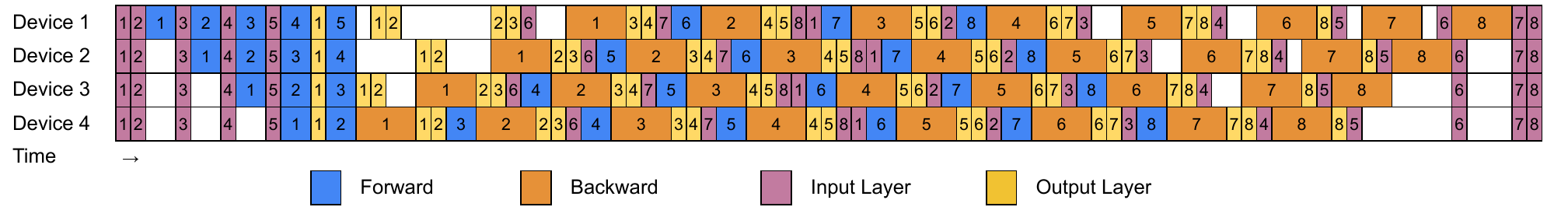}
    \caption{Full 1F1B schedules with Vocabulary Parallelism, corresponding to (a) Algorithm \ref{alg:async1} and (b) Algorithm \ref{alg:async2} respectively. Algorithm \ref{alg:async1} requires activation memory for $p+2$ microbatches while Algorithm \ref{alg:async2} only requires $p+1$, where $p$ is the number of devices.}
    \label{fig:full-async-1}
\end{figure*}

Considering the building block of the schedule, by inserting 2 or 1 intervals (for Algorithms \ref{alg:async1} and \ref{alg:async2} respectively) between the forward and backward pass of the last transformer layer, we can create space to schedule the output layer computation. Within the repeating interval in the building block, we can schedule output layer passes ($S$ and $T$) arbitrarily in each pipeline device. We show an example based on 1F1B in Figure \ref{fig:building-block}, where a \textit{one-forward-one-backward-one-output} pattern is strictly followed. The final 1F1B schedules are presented in Figure \ref{fig:full-async-1}, which is produced by uniformly repeating the building blocks. Additionally, the building block for \textit{V-Half} can be found in Appendix \ref{app:v_half}.


For the peak activation memory, as we insert at most 2 intervals to the lifespan, the peak activation memory only increases by at most 2 microbatches, which is a small constant overhead. This is a remarkable improvement compared to synchronous pipeline schedules, which multiplies the activation memory requirement by 1.5 (see Appendix \ref{app:interlaced_pipeline_memory}). Furthermore, the memory savings from balancing the vocabulary parameters outweighs the increase in activation memory.

Notably, as shown in Figure \ref{fig:building-block}, the activation memory increased in terms of microbatches is equivalent to the number of communication barriers, which motivates our optimization of communication barriers in Section \ref{section:sync}.


\section{Experiments} \label{section:expt}

We construct our experiments to verify: a) Our schedules with Vocabulary Parallelism can bring acceleration; b) Our methods can achieve a balanced memory usage when combined with memory-balanced schedules like \textit{V-Half}~\citep{controlable-mem}; c) The partitioning of vocabulary layers has a reasonable scaling factor compared to linear scaling.

\subsection{Implementation}

We implement the pipeline scheduler and the partitioned vocabulary layers based on the open-source Megatron-LM project \cite{megatron-lm}.

Scheduling under the assumption that backward takes twice time as forward might introduce unnecessary bubbles, especially when these values differ significantly. As a result, we profile the run time of the passes and schedule the $S$ and $T$ passes accordingly\footnote{The profiling verifies whether the backward pass runs approximately twice the time as the forward pass (unrelated to the vocabulary layer), in order to insert the vocabulary passes appropriately. The pipeline schedule generated will remain unchanged unless this ratio deviates by a certain threshold. In practice, we find that the difference is negligible in most transformer networks, and these differences would not change the pipeline schedule. Hence, this profiling could be viewed as optional.}.

We handle the communication groups in separate streams, allowing the communication barrier to overlap with the transformer layer passes. We map the CUDA streams to separate GPU work queues to achieve this overlapping\footnote{See \href{https://docs.nvidia.com/deploy/mps/index.html\#cuda-device-max-connections}{https://docs.nvidia.com/deploy/mps/index.html\#cuda-device-max-connections}.}. However, this would affect communication-computation overlap performance of tensor parallelism \cite{megatron-lm} as it relies on single work queues. To mitigate this problem, we set all model parallel communication groups to use high-priority streams. Additionally, both AllReduce and Reduce mentioned on Algorithm \ref{alg:async1} and \ref{alg:async2} are implemented as NCCL \citep{githubGitHubNVIDIAnccl} AllReduce to avoid imbalanced communication volume across devices. 

We also pad the vocabulary size to be a multiple of $2p$ to improve memory alignment in the vocabulary layers, where $p$ is the number of devices. In particular, we observe an approximate 8\% increase in performance if our method is applied to 24 devices with padded size 256032 (a multiple of 48), compared to the original value 256008.

Note that our method makes tying input and output embedding weights easier as the input and output embedding weights now have the same device placement and can use the shared weight tensor. This saves GPU memory and avoids the additional all-reduce to synchronize gradients. However, in all our experiments, we adapted the more difficult setting, untying the input and output embedding weights, since it is adapted by some open source models like Llama 3 \citep{dubey2024llama}.

\subsection{Setup} \label{section:expt-setup}

We compare the following methods implemented on the 1F1B schedule \cite{1f1b}.

\begin{itemize}
    \setlength{\itemsep}{0pt}
    \item Baseline: The na\"ive implementation in Megatron-LM. It distributes the transformer layers equally to all pipeline stages, while assigning the input and output layers to the first and last pipeline devices. This leads to highly imbalanced compute and memory.
    \item Redis: Redistributes the transformer layers across pipeline stages to balance out the computation as much as possible. We follow the derivation by \citeauthor{megatron-lm} \yrcite{megatron-lm} to estimate the number of floating point operations in each pipeline stage, and minimize the length of the longest stage.
    \item Vocab-1: Implements Vocabulary Parallelism with only forward phase optimization, as described in Algorithm \ref{alg:async1}.
    \item Vocab-2: On top of Vocab-1, applies backward phase optimization, as described in Algorithm \ref{alg:async2}.
    \item Interlaced: Our implementation of the fully synchronous interlaced pipeline proposed by \citeauthor{nnScaler} \yrcite{nnScaler}.
\end{itemize}

We experiment the implementations by pretraining GPT-like models of varying model and vocabulary sizes with up to 32 NVIDIA A100 SXM 80G GPUs distributed across 4 nodes, interconnected by a RoCE RDMA network. The running time of each iteration is recorded after several warm-up iterations. We compare the 5 methods under each fixed setting of model and vocabulary size, shown in Table \ref{table:exptsettings}.

Our experiments use pure pipeline parallelism to verify that our method improves pipeline parallelism as expected. Given that our method is orthogonal to tensor and data parallelism, the conclusions can be generalized to the production environment where pipeline parallelism is used together with tensor and data parallelism.

\begin{table}
\caption{Settings used in experiments on 1F1B schedule.}
\label{table:exptsettings}
\vskip 0.15in
\begin{center}
\begin{small}
\begin{sc}
\begin{tabular}{lccc}
\toprule
Pipelines (GPUs) & 8 & 16 & 32 \\
\midrule
Model Size & $\approx$ 4B & $\approx$ 10B & $\approx$ 21B \\
Layers & 32 & 48 & 64 \\
Attention Heads & 24 & 32 & 40 \\
Hidden Size & 3072 & 4096 & 5120 \\
Sequence Length & \multicolumn{3}{c}{2048 / 4096} \\
Microbatch Size & \multicolumn{3}{c}{1} \\
Number of Microbatches & \multicolumn{3}{c}{128} \\
Vocabulary Size & \multicolumn{3}{c}{32k / 64k / 128k / 256k} \\
\bottomrule
\end{tabular}
\end{sc}
\end{small}
\end{center}
\vskip -0.1in
\end{table}

\begin{figure}[t]
    \centering
    \includegraphics[width=\linewidth]{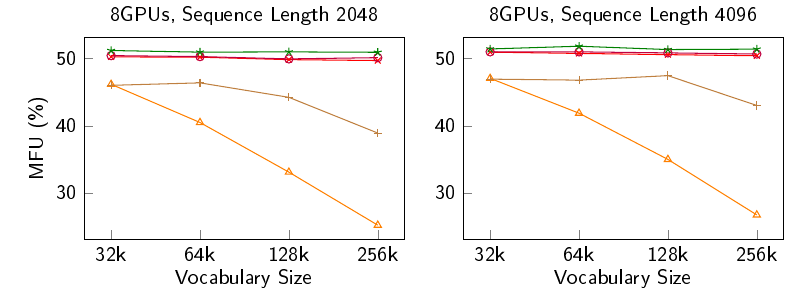}
    \includegraphics[width=\linewidth]{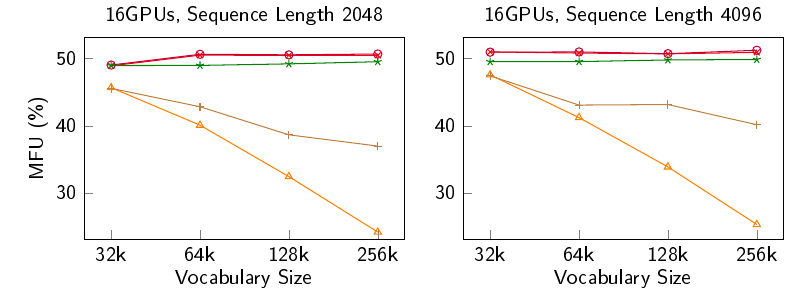}
    \includegraphics[width=\linewidth]{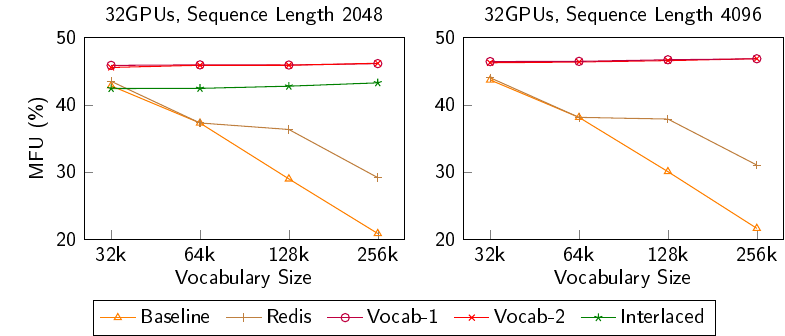}
    \caption{Throughput of different methods on 1F1B. Interlaced OOMs when training with 32 GPUs and sequence length 4096.}
    \label{fig:1f1b-pp-mfu}
\end{figure}

\begin{figure}[t]
    \centering
    \includegraphics[width=\linewidth]{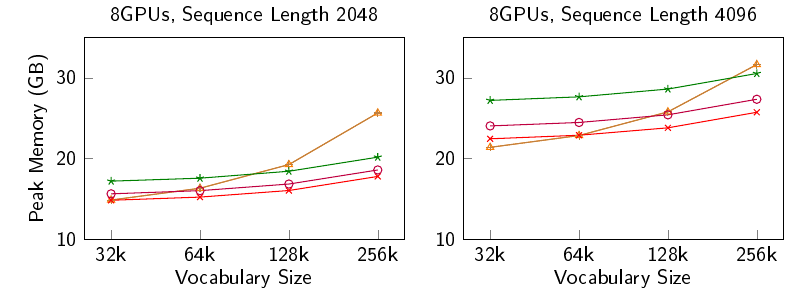}
    \includegraphics[width=\linewidth]{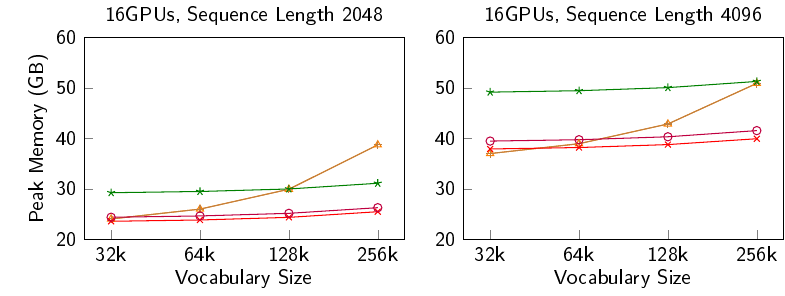}
    \includegraphics[width=\linewidth]{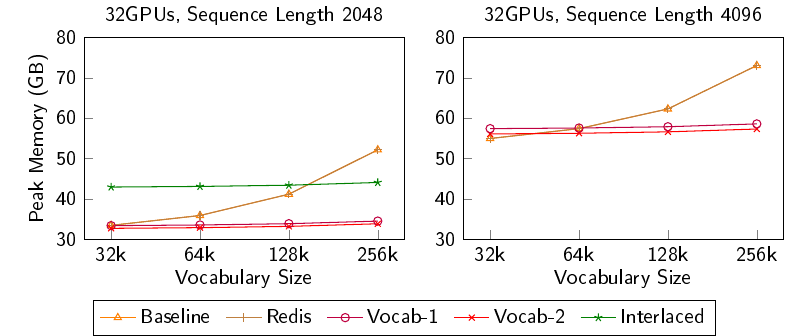}
    \caption{Peak memory of different methods on 1F1B}
    \label{fig:1f1b-pp-mem}
\end{figure}

\begin{figure}[t]
    \centering
    \includegraphics[width=\linewidth]{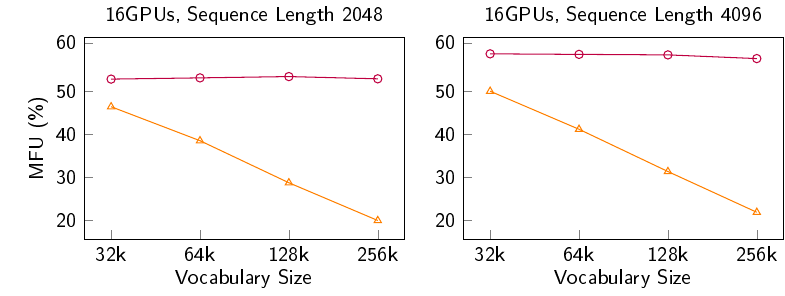}
    \includegraphics[width=\linewidth]{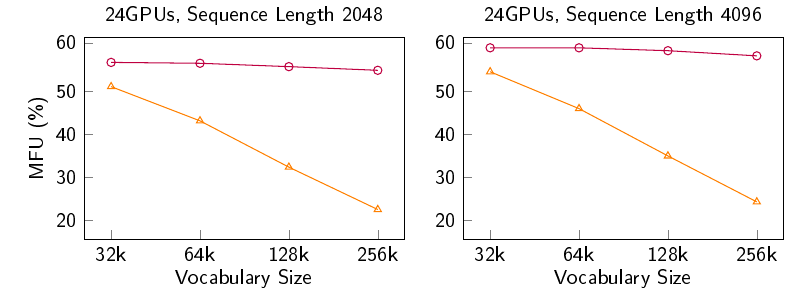}
    \includegraphics[width=\linewidth]{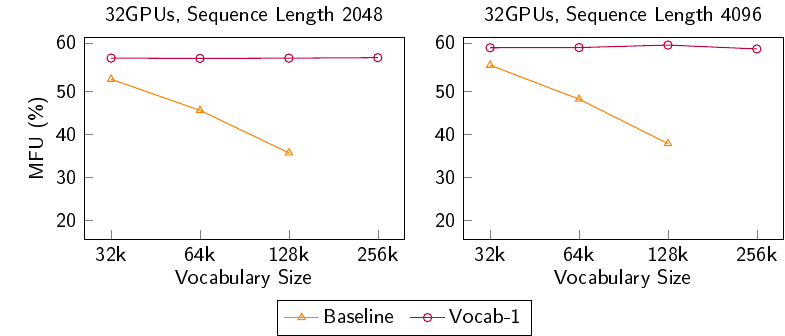}
    \caption{Throughput of different methods on \textit{V-Half}. Baseline OOMs when training with 32 GPUs and vocabulary size 256k.}
    \label{fig:vhalf-pp-mfu}
\end{figure}

\begin{figure}[t]
    \centering
    \includegraphics[width=\linewidth]{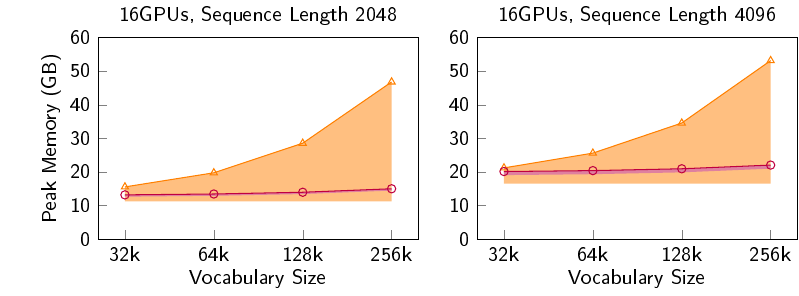}
    \includegraphics[width=\linewidth]{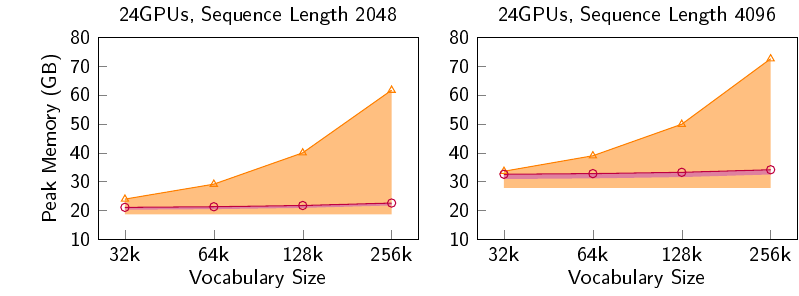}
    \includegraphics[width=\linewidth]{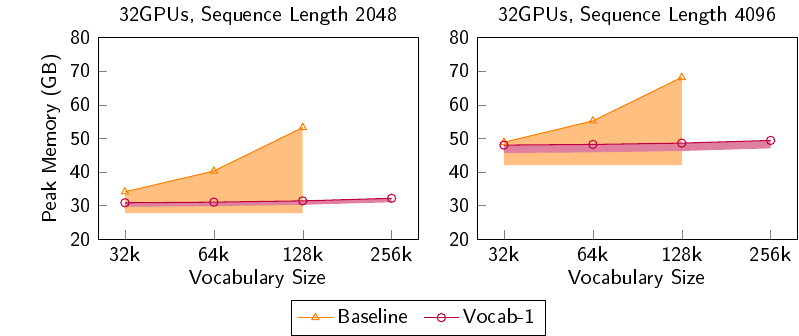}
    \caption{Peak memory of different methods on \textit{V-Half}. The shaded area denotes the range of maximum allocated memory for all devices.}
    \label{fig:vhalf-pp-mem}
\end{figure}

\subsection{Comparison of Methods} \label{section:compare1f1b}

We present comparisons of the throughput measured in FLOPs utilization (MFU) and peak memory in Figures \ref{fig:1f1b-pp-mfu} and \ref{fig:1f1b-pp-mem}, respectively. As shown in the figures, the layer redistribution approach suffers from a significant performance degradation of 8\% to 33\% for large vocabulary sizes, since the output layer alone already has a higher computation cost than that in the other pipeline devices. Its performance is also highly dependent on the model configuration, or more specifically, the ratio of compute between the vocabulary layer and transformer layers. For example, there is a 9.7\% drop in MFU when increasing the vocabulary size from 64k to 128k for the 10B model with sequence length 2048, but that is not observed with sequence length 4096. In contrast, the Vocab and Interlaced approaches have a consistent MFU when scaling up the vocabulary sizes. Vocabulary Parallelism outperforms the interlaced pipeline under a multi-node setup, due to its overlapped communication. For the 21B model, Vocabulary Parallelism outperforms the interlaced pipeline by 6.7\% to 8.2\% in MFU.

For peak memory usage, the na\"ive implementation and layer redistribution approaches have an imbalanced parameter memory, leading to high peak memory for large vocabulary sizes. Although the Vocabulary Parallelism methods require extra activation memory, it is effectively negligible when we scale up the pipeline parallel size. However, the interlaced pipeline requires 1.5 times activation memory compared to 1F1B. This resulted in out-of-memory when training the 21B model with sequence length 4096.

\subsection{Memory-Balanced Schedule} \label{section:comparevhalf}

We show that our method can achieve a balanced memory usage by applying Vocab-1 on the \textit{V-Half} schedule \cite{controlable-mem}, a memory-balanced schedule. The implementation is based on the open-sourced \textit{V-Half} implementation by \citeauthor{controlable-mem} (\citeyear{controlable-mem}). To support division into virtual pipeline chunks, we adopt different configurations in the experiments, as shown in Table \ref{vhalf-settings}.

\begin{table}[h!]
\caption{Settings used in experiments on \textit{V-Half} schedule.}
\label{vhalf-settings}
\vskip 0.15in
\begin{center}
\begin{small}
\begin{sc}
\begin{tabular}{lccc}
\toprule
Pipelines (GPUs) & 16 & 24 & 32 \\
\midrule
Model Size & $\approx$ 7B & $\approx$ 16B & $\approx$ 30B \\
Layers & 32 & 48 & 64 \\
Attention Heads & 32 & 40 & 48 \\
Hidden Size & 4096 & 5120 & 6144 \\
Sequence Length & \multicolumn{3}{c}{2048 / 4096} \\
Microbatch Size & \multicolumn{3}{c}{1} \\
Number of Microbatches & \multicolumn{3}{c}{128} \\
Vocabulary Size & \multicolumn{3}{c}{32k / 64k / 128k / 256k} \\
\bottomrule
\end{tabular}
\end{sc}
\end{small}
\end{center}
\vskip -0.1in
\end{table}

\pagebreak

We compare the na\"ive \textit{V-Half} schedule implementation and that incorporated with Vocab-1. The throughput and peak memory for each pipeline device are shown in Figures \ref{fig:vhalf-pp-mfu} and \ref{fig:vhalf-pp-mem} respectively. The na\"ive implementation resulted in out-of-memory in cases with 32 GPUs and a 256k vocabulary size.

Similar to the previous experiments, the baseline suffers from a huge performance drop when we increase the vocabulary size, while Vocab-1 maintains a steady MFU, consistently outperforming the baseline by 7.2\% to 143\%. Besides, the baseline has a significant memory imbalance across pipeline devices with up to 45GB difference, while Vocab-1 achieves a balanced memory usage across pipeline devices. Although the first pipeline device still holds slightly more parameters due to positional and token type embedding, the extra memory required is a small constant. In our experiments, this is less than 2.5GB.

\subsection{Scaling Analysis of Vocabulary Layers} \label{section:overhead}

We analyze the scalability of vocabulary layers in our Vocabulary Parallelism implementation. Using a vocabulary size of 256k, we measure the average throughput of all $S$ and $T$ passes across all devices in our implementation. We compare this with the ideal scenario where the vocabulary layers linearly scale (i.e. $p$ times of the original throughput when distributed to $p$ devices). 

The output layers for Vocab-1 and Vocab-2 are considered separately. The time used for the communications is not included since it overlaps with other computation. The results are shown in Table \ref{fig:overhead}.

\begin{table}[H]
\caption{The scaling factor of vocabulary layer computation relative to linear scaling on sequence lengths 2048 and 4096.}
\label{fig:overhead}
\vskip 0.15in
\begin{center}
\begin{small}
\begin{sc}
\begin{tabular}{ccccc}
\toprule
Seq & Layer & 8GPU & 16GPU & 32GPU \\
\midrule
\multirow{3}{*}{2048} & Output-Vocab-1 & 91.29\% & 84.22\% & 80.59\% \\
 & Output-Vocab-2 & 86.72\% & 79.84\% & 75.93\% \\
 & Input & 39.99\% & 28.85\% & 15.18\% \\
\midrule
\multirow{3}{*}{4096} & Output-Vocab-1 & 93.21\% & 88.02\% & 85.24\% \\
 & Output-Vocab-2 & 88.36\% & 83.42\% & 79.66\% \\
 & Input & 27.69\% & 15.52\% & 8.35\% \\
\bottomrule
\end{tabular}
\end{sc}
\end{small}
\end{center}
\vskip -0.1in
\end{table}

Parallelizing the vocabulary layers comes with some computation overhead, which can be attributed to two causes. Firstly, partitioning the vocabulary layers will reduce the model FLOPs utilization (MFU) of GPU kernels as the operations are smaller and hence less parallelized. Secondly, this brings extra computation, especially for the input layer where all devices have to construct the output tensor, whose size is independent of the size of the vocabulary partition. However, it's worth noting that both input and output still only take a small portion of the computation of the entire model after being partitioned.


\section{Conclusion and Future Work}

In this work, we identified the problem that when training LLMs with pipeline parallelism,  vocabulary layers  causes non-negligible imbalance for both compute and memory. Existing methods either fails to achieve a balance or introduce significant performance overhead to the original pipeline schedule. To address this issue, we proposed Vocabulary Parallelism, a method that partitions vocabulary layers evenly to pipeline devices and integrates them into existing pipeline schedules. Our method achieves compute and memory balance for the vocabulary layers. As a result, experiments shows that it improves the throughput by up to 51\% while also reduces peak memory consumption compared to existing methods.

Although our implementation of the vocabulary layers are pure python-based, we find that similar optimizations to Algorithm \ref{alg:async2} opens an opportunity of fusing the forward and backward pass in CUDA kernels to avoid writes/reads of the softmax results, which can be huge in long-context large-vocabulary settings, to main memory, similar to the rationale of FlashAttention \citep{dao2022flashattention}. Also, while our work focuses on the imbalanced vocabulary layers for pure text-based LLMs, we believe the embedding layers for multimodal LLMs suffer from the same problem and can be further explored. 

\newpage

\bibliography{main}
\bibliographystyle{mlsys2024}


\appendix

\vfill\null
\pagebreak

\section{Quantitative Analysis of Vocabulary Layers} \label{app:quantitative_analysis}

Following the calculations of \citeauthor{megatron-lm} \yrcite{megatron-lm} and neglecting insignificant terms, we present the computational and memory expenses in relation to a single transformer layer in Table \ref{tab:flopsmem}. In this table we denote the microbatch size as $b$, sequence length as $s$, hidden dimension as $h$ and vocabulary size as $V$. It is worth noting that the activation memory is excluded from this analysis, as it typically has a transient nature for vocabulary layers.
\begin{table}[H]
\caption{Compute and memory cost of vocabulary and transformer layers}
\label{tab:flopsmem}
\vskip 0.15in
\begin{center}
\begin{small}
\begin{sc}
\begin{tabular}{lcc}
\toprule
Layer Type & Compute FLOPs & Param Memory \\
\midrule
Transformer & $bsh(72h+12s)$ & $24h^2$ \\
Input & $3bsh$ & $2hV$ \\
Output & $6bshV$ & $2hV$ \\

\bottomrule
\end{tabular}
\end{sc}
\end{small}
\end{center}
\vskip -0.1in
\end{table}

\section{More Analysis of Interlaced Pipeline}

\subsection{Memory Analysis} \label{app:interlaced_pipeline_memory}

\begin{figure}[H]
    \centering
    \begin{subfigure}[b]{\linewidth}
        \centering
        \includegraphics[width=\linewidth]{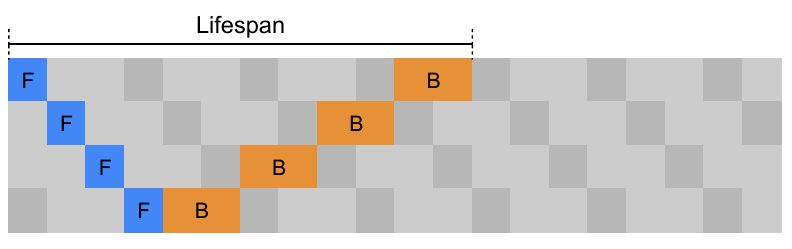}
        \caption{Building block of 1F1B}
        \label{fig:1f1b-grid}
    \end{subfigure}
    \begin{subfigure}[b]{\linewidth}
        \centering
        \includegraphics[width=\linewidth]{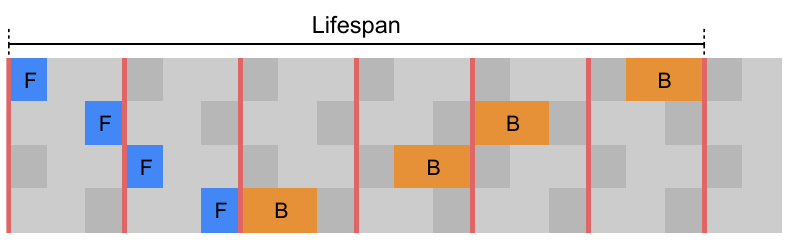}
        \caption{Building block of interlaced pipeline parallel. The red vertical lines indicate synchronization introduced by TP of vocabulary layers.}
        \label{fig:interlace-grid}
    \end{subfigure}
    \caption{Comparison between building blocks of 1F1B and Interlaced PP.}
    \label{fig:compare-grid}
\end{figure}

One concern of interlaced pipeline is that the peak activation memory of 1F1B is raised to 1.5 times of its original value. This increase in memory consumption can be analyzed using the framework introduced in \citeauthor{controlable-mem} \yrcite{controlable-mem}. As shown in Figure \ref{fig:compare-grid}, the interlaced schedule enlarges the lifespan of 1F1B's building block from $3p$ to approximately $4.5p$ where $p$ is the number of devices, resulting in 1.5x peak activation memory consumption.

\subsection{Overhead of Tensor Parallel Communication} \label{app:interlaced_pipeline_sync}

In practice, tensor parallel is only used for intra-node parallelism due to its high communication volume. The interlaced pipeline has introduced a tensor parallel style parallelization for the vocabulary layers, which creates additional pipeline bubbles for each microbatch during the tensor parallel communication. To quantify the size of the bubbles, we conduct an ablation study by training a 21.5B model using 32 GPUs. We remove the synchronous all-reduce communications in the vocabulary layers, and measure the speedup in end-to-end iteration time. Note that the all-reduce communications that are overlapped with the computation are still kept.

By removing the synchronous all-reduce communications, the end-to-end iteration time improved by 10.95\%. This shows that the synchronous all-reduce communications contributed to approximately 11\% of the idle time when training using an interlaced pipeline. We conclude that the interlaced pipeline is undesirable for multi-node training.

\section{Vocabulary Parallelism for The Input Layer} \label{appendix:input-layer}

While the output layers involve complex dependencies and communications, input layer computation can be completed independently before and after the transformer layer passes. The only required communications are an all-reduce communication after the forward pass, and a broadcast communication before the backward pass. These communications can be overlapped with the transformer layer computation, and can be scheduled well-ahead or after.

We schedule the input layer passes as follows:
\begin{itemize}
    \item During the warm-up forward passes, we insert the input layer forward pass one microbatch before the first transformer layer forward pass. This allows time for gathering the input layer outputs.

    \item In the stable phase, we piggyback the input layer forward pass with the output layer passes, scheduled as least one repeating interval beforehand. Similarly, the input layer backward passes are piggybacked at least one repeating interval afterwards, allowing enough time to broadcast the output gradient to all devices.

    \item During the cool-down backward passes, we insert the input layer backward pass one microbatch after the last transformer layer backward pass.
\end{itemize}

This schedule ensures that each device is holding the input layer outputs for at most two microbatches at any instant, reducing the memory pressure.

\section{\textit{V-Half} Pipeline Scheduling} \label{app:v_half}

\begin{figure}[H]
    \centering
    \includegraphics[width=\linewidth]{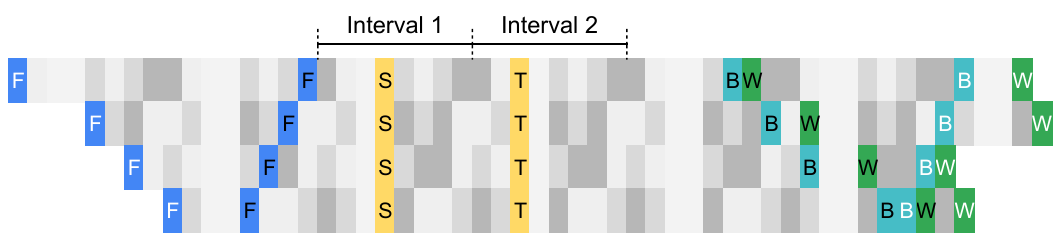}
    \caption{Modified building block for the \textit{V-Half} schedule.}
    \label{fig:vhalf-building-block}
\end{figure}

Following the scheduling methodology in section \ref{section:scheduling-examples}, we show the building block for the \textit{V-Half} schedule in Figure \ref{fig:vhalf-building-block}.

\section{Correctness Evaluation}

Our implementation is based on the open-source Megatron-LM project \cite{megatron-lm}. We compare the convergence curves of our implementation to that of the original Megatron-LM codebase to verify our implementation's correctness. The configurations follow the 4B model in section \ref{section:expt-setup} with a vocabulary size of 256K, trained with 8 GPUs. Additionally, we verify that our implementation also works correctly with tensor parallelism, by using a pipeline parallel size of 4 and a tensor parallel size of 2.

\begin{figure}[h!]
    \centering
    \includegraphics[width=\linewidth]{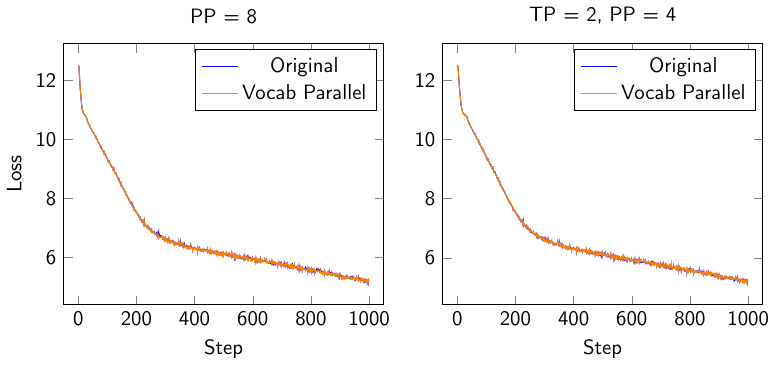}
    \caption{Convergence curves of our implementation against the original Megatron-LM codebase}
    \label{fig:convergence}
\end{figure}

The convergence curves are shown in Figure \ref{fig:convergence}. It is shown that our implementation maintains correctness, albeit with some small numerical differences.

\section{Detailed Experiment Data}

For Sections \ref{section:compare1f1b} and \ref{section:comparevhalf}, we present the detailed experimental data in Tables \ref{table:compare1f1b} and \ref{table:comparevhalf} respectively. The following metrics are computed:

\begin{itemize}
    \item MFU: The FLOPs utilization of the training system. We follow \citeauthor{megatron-lm} (\citeyear{megatron-lm})'s derivation to compute the FLOPs of the model.
    \item Peak Memory: The maximum peak memory across all devices.
\end{itemize}

\begin{table*}[t]
\caption{Comparison of Methods on 1F1B.}
\label{table:compare1f1b}
\vskip 0.15in
\begin{center}
\begin{small}
\begin{sc}
\begin{tabular}{c|c|cccc|cccc}
\toprule
\multirow{2}{*}{Setup} & \multirow{2}{*}{Method} & \multicolumn{4}{|c}{MFU (\%)} & \multicolumn{4}{|c}{Peak Memory (GB)} \\
& & 32k & 64k & 128k & 256k & 32k & 64k & 128k & 256k \\
\midrule
\multirow{5}{*}{8GPU, Seq Length 2048} & Baseline & 46.16 & 40.48 & 33.11 & 25.23 & 14.86 & 16.32 & 19.25 & 25.64 \\
& Redis & 46.01 & 46.37 & 44.22 & 38.91 & 14.86 & 16.32 & 19.25 & 25.64\\
& Vocab-1 & 50.42 & 50.28 & 49.93 & 50.12 & 15.63 & 16.02 & 16.84 & 18.59 \\
& Vocab-2 & 50.23 & 50.18 & 49.82 & 49.69 & \textbf{14.83} & \textbf{15.23} & \textbf{16.04} & \textbf{17.78} \\
& Interlaced & \textbf{51.18} & \textbf{50.94} & \textbf{50.97} & \textbf{50.92} & 17.20 & 17.57 & 18.43 & 20.17 \\
\midrule
\multirow{5}{*}{8GPU, Seq Length 4096} & Baseline & 47.05 & 41.87 & 35.00 & 26.75 & \textbf{21.39} & \textbf{22.85} & 25.78 & 31.64 \\
& Redis & 46.93 & 46.78 & 47.44 & 43.01 & \textbf{21.39} & \textbf{22.85} & 25.78 & 31.64 \\
& Vocab-1 & 50.98 & 50.98 & 50.83 & 50.66 & 24.04 & 24.47 & 25.41 & 27.34 \\
& Vocab-2 & 50.93 & 50.75 & 50.56 & 50.40 & 22.44 & 22.89 & \textbf{23.80} & \textbf{25.73} \\
& Interlaced & \textbf{51.41} & \textbf{51.82} & \textbf{51.32} & \textbf{51.38} & 27.20 & 27.64 & 28.60 & 30.53 \\
\midrule
\multirow{5}{*}{16GPU, Seq Length 2048} & Baseline & 45.66 & 40.09 & 32.44 & 24.21 & 24.03 & 25.98 & 29.92 & 38.71 \\
& Redis & 45.56 & 42.82 & 38.65 & 36.98 & 24.03 & 25.98 & 29.92 & 38.71 \\
& Vocab-1 & \textbf{49.02} & \textbf{50.62} & \textbf{50.54} & \textbf{50.66} & 24.37 & 24.63 & 25.14 & 26.26 \\
& Vocab-2 & 48.90 & 50.49 & 50.46 & 50.46 & \textbf{23.57} & \textbf{23.83} & \textbf{24.35} & \textbf{25.47} \\
& Interlaced & 48.94 & 48.97 & 49.19 & 49.52 & 29.23 & 29.47 & 29.97 & 31.10 \\
\midrule
\multirow{5}{*}{16GPU, Seq Length 4096} & Baseline & 47.56 & 41.21 & 33.88 & 25.33 & \textbf{36.99} & 38.94 & 42.85 & 50.90 \\
& Redis & 47.41 & 43.07 & 43.15 & 40.15 & \textbf{36.99} & 38.94 & 42.85 & 50.90 \\
& Vocab-1 & 50.93 & \textbf{50.97} & \textbf{50.71} & \textbf{51.22} & 39.46 & 39.73 & 40.31 & 41.53 \\
& Vocab-2 & \textbf{50.97} & 50.80 & 50.68 & 50.90 & 37.89 & \textbf{38.18} & \textbf{38.77} & \textbf{39.92} \\
& Interlaced & 49.52 & 49.53 & 49.77 & 49.84 & 49.16 & 49.44 & 50.05 & 51.28 \\
\midrule
\multirow{5}{*}{32GPU, Seq Length 2048} & Baseline & 42.81 & 37.28 & 28.97 & 20.86 & 33.45 & 35.89 & 41.17 & 52.16 \\
& Redis & 43.48 & 37.29 & 36.32 & 29.16 & 33.45 & 35.89 & 41.17 & 52.16 \\
& Vocab-1 & \textbf{45.85} & \textbf{45.92} & \textbf{45.90} & 46.11 & 33.38 & 33.55 & 33.86 & 34.51 \\
& Vocab-2 & 45.54 & 45.86 & 45.86 & \textbf{46.16} & \textbf{32.72} & \textbf{32.88} & \textbf{33.20} & \textbf{33.84} \\
& Interlaced & 42.40 & 42.43 & 42.75 & 43.25 & 42.94 & 43.09 & 43.40 & 44.07 \\
\midrule
\multirow{5}{*}{32GPU, Seq Length 4096} & Baseline & 43.68 & 38.11 & 30.05 & 21.63 & \textbf{54.97} & 57.41 & 62.29 & 73.05 \\
& Redis & 44.01 & 38.12 & 37.87 & 31.03 & \textbf{54.97} & 57.41 & 62.29 & 73.05 \\
& Vocab-1 & \textbf{46.41} & \textbf{46.44} & \textbf{46.68} & 46.83 & 57.41 & 57.56 & 57.88 & 58.58 \\
& Vocab-2 & 46.23 & 46.35 & 46.55 & \textbf{46.84} & 56.09 & \textbf{56.26} & \textbf{56.61} & \textbf{57.31} \\
& Interlaced & - & - & - & - & - & - & - & - \\
\bottomrule
\end{tabular}
\end{sc}
\end{small}
\end{center}
\vskip -0.1in
\end{table*}

\begin{table*}[t]
\caption{Comparison of Methods on \textit{V-Half}.}
\label{table:comparevhalf}
\vskip 0.15in
\begin{center}
\begin{small}
\begin{sc}
\begin{tabular}{c|c|cccc|cccc}
\toprule
\multirow{2}{*}{Setup} & \multirow{2}{*}{Method} & \multicolumn{4}{|c}{MFU (\%)} & \multicolumn{4}{|c}{Peak Memory (GB)} \\
& & 32k & 64k & 128k & 256k & 32k & 64k & 128k & 256k \\
\midrule
\multirow{2}{*}{16GPU, Seq Length 2048} & Baseline & 46.41 & 38.52 & 28.75 & 19.99 & 15.57 & 19.77 & 28.55 & 46.77 \\
& Vocab-1 & \textbf{52.82} & \textbf{53.11} & \textbf{53.41} & \textbf{52.89} & \textbf{13.20} & \textbf{13.46} & \textbf{13.98} & \textbf{15.02} \\
\midrule
\multirow{2}{*}{16GPU, Seq Length 4096} & Baseline & 50.01 & 41.17 & 31.36 & 21.90 & 21.22 & 25.61 & 34.56 & 53.11 \\
& Vocab-1 & \textbf{58.69} & \textbf{58.56} & \textbf{58.44} & \textbf{57.59} & \textbf{20.14} & \textbf{20.41} & \textbf{20.96} & \textbf{22.06} \\
\midrule
\multirow{2}{*}{24GPU, Seq Length 2048} & Baseline & 51.07 & 43.13 & 32.38 & 22.54 & 23.94 & 29.12 & 39.98 & 61.71 \\
& Vocab-1 & \textbf{56.70} & \textbf{56.50} & \textbf{55.72} & \textbf{54.86} & \textbf{21.08} & \textbf{21.29} & \textbf{21.72} & \textbf{22.57} \\
\midrule
\multirow{2}{*}{24GPU, Seq Length 4096} & Baseline & 54.53 & 45.96 & 34.99 & 24.31 & 33.60 & 38.97 & 49.90 & 72.60 \\
& Vocab-1 & \textbf{60.09} & \textbf{60.09} & \textbf{59.42} & \textbf{58.22} & \textbf{32.55} & \textbf{32.78} & \textbf{33.22} & \textbf{34.12} \\
\midrule
\multirow{2}{*}{32GPU, Seq Length 2048} & Baseline & 52.80 & 45.56 & 35.69 & - & 34.11 & 40.28 & 53.22 & - \\
& Vocab-1 & \textbf{57.70} & \textbf{57.62} & \textbf{57.69} & \textbf{57.80} & \textbf{30.85} & \textbf{31.04} & \textbf{31.42} & \textbf{32.18} \\
\midrule
\multirow{2}{*}{32GPU, Seq Length 4096} & Baseline & 56.06 & 48.17 & 37.85 & - & 48.84 & 55.19 & 68.12 & - \\
& Vocab-1 & \textbf{60.10} & \textbf{60.14} & \textbf{60.72} & \textbf{59.82} & \textbf{47.99} & \textbf{48.19} & \textbf{48.59} & \textbf{49.38} \\
\bottomrule
\end{tabular}
\end{sc}
\end{small}
\end{center}
\vskip -0.1in
\end{table*}

\clearpage
\section{Artifact Appendix}

\subsection{Abstract}

This section will outline the setup and experimental workflow of \textbf{Balancing Pipeline Parallelism with Vocabulary Parallelism} conducted on a single server equipped with 8 A100 GPUs.


\subsection{Artifact check-list (meta-information)}

\begin{minipage}[t]{0.48\textwidth}
{\small
\begin{itemize}[itemsep=2pt,parsep=2pt]
  \item {\bf Algorithm: } Vocabulary Parallelism, Online Softmax
  \item {\bf Program: } Not used
  \item {\bf Compilation: } Nvidia nvcc version 12.4, already in the container
  \item {\bf Transformations: } Not used
  \item {\bf Binary: } will be compiled on a target platform
  \item {\bf Data set: } Customized C4 hosted in Huggingface
  \item {\bf Binary: } will be compiled on a target platform.
  \item {\bf Run-time environment: } Ubuntu 20.04.6. Needs docker. Requires root.
  \item {\bf Hardware: } Server with 8 Nvidia A100 GPUs 80GB HBM.
  \item {\bf Run-time state: } Server is idle.
  \item {\bf Execution: } Takes at least 4 hours to complete.
  \item {\bf Metrics: } Peak Memory, MFU
  \item {\bf Output: } The data printed on console. The output of Quick Experiment is the key result.
  \item {\bf Experiments: } Elaborated in the Installation and Experiment workflow sections.
  \item {\bf How much disk space required (approximately)?: } 30 GB
  \item {\bf How much time is needed to prepare workflow (approximately)?: } 30 minutes
  \item {\bf How much time is needed to complete experiments (approximately)?: } 4 hours
  \item {\bf Publicly available?: } Yes
  \item {\bf Code licenses (if publicly available)?: } Apache License
  \item {\bf Data licenses (if publicly available)?: } odc-by
  \item {\bf Workflow framework used?: } No
  \item {\bf Archived (provide DOI)?: }
\end{itemize}
}
\end{minipage}
\hfill

\subsection{Description}

This experiment consists of 2 parts:

\begin{itemize}
    \item \textbf{Quick Experiment} to quickly verify the result on a specific case.
    \item \textbf{Full Experiment} to run all cases on an 8-GPU server.
\end{itemize}

\noindent The \textbf{Full Experiment} employs the settings in table \ref{table:exptsettings_artifact} same as the paper. The \textbf{Quick Experiment} focuses on a specific case where the sequence length is 4096 and the vocabulary size is 256k. We use Megatron-LM on which all methods are implemented to run training benchmarks. 

\begin{table}[H]
\caption{Artifact Settings used in experiments on 1F1B schedule.}
\label{table:exptsettings_artifact}
\vskip 0.15in
\begin{center}
\begin{small}
\begin{sc}
\begin{tabular}{lccc}
\toprule
Pipelines (GPUs) & 8 \\
\midrule
Model Size & $\approx$ 4B \\
Layers & 32 \\
Attention Heads & 24 \\
Hidden Size & 3072\\
Sequence Length & \multicolumn{1}{c}{2048 / 4096} \\
Microbatch Size & \multicolumn{1}{c}{1} \\
Number of Microbatches & \multicolumn{1}{c}{128} \\
Vocabulary Size & \multicolumn{1}{c}{32k / 64k / 128k / 256k} \\
\bottomrule
\end{tabular}
\end{sc}
\end{small}
\end{center}
\vskip -0.1in
\end{table}

\subsubsection{How delivered}

 The code locates on
 \href{https://github.com/sail-sg/VocabularyParallelism}{\url{https://github.com/sail-sg/VocabularyParallelism}}. The \href{https://huggingface.co/datasets/mtyeung/vocab_parallel_sample_dataset}{dataset} will be automatically downloaded in experiment scripts.

\subsubsection{Hardware dependencies}

The tests should be conducted in a server with 8 Nvidia A100 GPUs, 80GB HBM.

\subsubsection{Software dependencies}
\begin{itemize}
    \item \textbf{CUDA Driver Version}: 535.54.03
    \item \textbf{CUDA Version} 12.2
    \item \textbf{Docker}
    \item \textbf{NVIDIA Container Toolkit}
\end{itemize}

\subsubsection{Data sets}
The dataset is customized from C4 hosted in Huggingface \newline
\href{https://huggingface.co/datasets/mtyeung/vocab\_parallel\_sample\_dataset}{\url{https://huggingface.co/datasets/mtyeung/vocab\_parallel\_sample\_dataset}}
supporting various sequence length. It will be automatically downloaded in experiment scripts.

\vspace{15mm}

\subsection{Installation}

\noindent Run a container:
\begin{lstlisting}[language=bash]
docker run --name vocab_torch24 \
    --network=host -d \
    --runtime=nvidia --gpus all \
    --ipc=host --ulimit memlock=-1 --ulimit stack=67108864 \
    --privileged=true \
    nvcr.io/nvidia/pytorch:24.03-py3 sleep infinity
\end{lstlisting}

\noindent Get inside the container, and clone the codes:
\begin{lstlisting}[language=bash]
# Enter the container
docker exec -it vocab_torch24 bash
# Clone the codes
git clone https://github.com/sail-sg/VocabularyParallelism.git
cd VocabularyParallelism
\end{lstlisting}

 \noindent Note that all the following commands should be run inside the \texttt{VocabularyParallelism} directory.

\subsection{Experiment workflow}

\subsubsection{Quick Experiment}
The quick experiment runs all the methods (\textit{baseline}, \textit{redis}, \textit{interlaced}, \textit{vocab-1}, \textit{vocab-2}) on a specific setting in the paper:
\begin{itemize}[itemsep=1pt,parsep=1pt,topsep=0pt]
    \item Sequence Length: 4096
    \item Vocabulary Size: 256k
\end{itemize}

\noindent The experiment will show 2 key results:
\begin{itemize}[itemsep=1pt,parsep=1pt,topsep=0pt]
    \item \textbf{Peak Memory}
    \item \textbf{MFU}
\end{itemize}

\noindent Run all the methods one by one:
\begin{lstlisting}[language=bash]
bash artifact/quick_exp.sh run baseline
bash artifact/quick_exp.sh run redis
bash artifact/quick_exp.sh run interlaced
bash artifact/quick_exp.sh run vocab-1
bash artifact/quick_exp.sh run vocab-2
\end{lstlisting}

\noindent This will automatically download the dataset from huggingface and run the training experiments. The log containing the result will locate in \texttt{quick-logs/<method>/stdout.log}. Each method should take around 6 minutes to complete.

\noindent Then run this to collect the results:
\begin{lstlisting}[language=bash]
bash artifact/quick_exp.sh show-result
\end{lstlisting}

\subsubsection{Full Experiment}
This will run all experiments on a single server with 8 A100 GPUs.

\noindent The whole experiment will take around 3 hours to complete.
\begin{lstlisting}[language=bash]
bash artifact/full_exp.sh artifact/exp_one_host.csv
\end{lstlisting}

\noindent Print results:
\begin{lstlisting}[language=bash]
python artifact/show_result_full_exp.py
\end{lstlisting}

\subsection{Evaluation and expected result}

The output of \textbf{Quick Experiment} should show similar result as
\newline \begin{small}
\href{https://github.com/sail-sg/VocabularyParallelism/blob/main/artifact/example-results/quick-exp.txt}{\url{https://github.com/sail-sg/VocabularyParallelism/blob/main/artifact/example-results/quick-exp.txt}}
\end{small}

\noindent The expected output of \textbf{Full Experiment} is located in
\newline \begin{small}
\href{https://github.com/sail-sg/VocabularyParallelism/blob/main/artifact/example-results/full-exp.txt}{\url{https://github.com/sail-sg/VocabularyParallelism/blob/main/artifact/example-results/full-exp.txt}}
\end{small}

\noindent The result should also roughly match the 2 rows in \textbf{Table 5. Comparison of Methods on 1F1B} in the paper:

\begin{itemize}[itemsep=1pt,parsep=1pt,topsep=0pt]
    \item 8GPU, SEQ LENGTH 2048
    \item 8GPU, SEQ LENGTH 4096
\end{itemize}

\noindent The result shows that the throughput and peak memory of vocab-1 and vocab-2 are significantly better than baseline and redistribution.
The throughput of interlaced is slightly better than vocab-1 and vocab-2. But the peak memory of interlaced is worse than our approach.

\subsection{Experiment customization}

User can customize the settings by changing the scripts quick\_exp.sh, full\_exp.sh, show\_result\_full\_exp.py under under \href{https://github.com/sail-sg/VocabularyParallelism/blob/main/artifact/}{artifact/}.





\end{document}